\documentstyle[11pt]{article}
\def\be{\begin{equation}}
\def\ee{\end{equation}}
\def\bea{\begin{eqnarray}}
\def\eea{\end{eqnarray}}
\def\l{\label}
\def\c{\cite}

\begin{document}
\begin{titlepage}

\vspace{1in}

\begin{center}
\Large
{\bf The Andante Regime of Scalar Field Dynamics}

\vspace{.35in}

\normalsize

 James E. Lidsey$^{\ast ,}$\footnote{Address from $1$st October 1994:
Astronomy Unit, School of Mathematical Sciences, Mile End Road, LONDON
E1 4NS, U.K.} \& I. Waga$^{\ast ,\diamond}$

\normalsize
\vspace{.7cm}

{\em $^{\ast}$NASA/Fermilab Astrophysics Center,  \\
Fermi National Accelerator Laboratory, Batavia, IL 60510}

\vspace{.35cm}

{\em $^{\diamond}$Universidade Federal do Rio de Janeiro, Instituto de
F\'isica, \\ Rio de Janeiro - RJ - Brasil -21943}

\end{center}

\vspace{.7in}

\baselineskip=24pt
\begin{abstract}
\noindent

The andante regime of scalar field dynamics in the chaotic
inflationary Universe is defined as the epoch when the field is
rolling moderately slowly down its interaction potential, but at such
a rate that first-order corrections to the slow-roll approximation
become important. These conditions should apply towards the end of
inflation as the field approaches the global minimum of the potential.
Solutions to the Einstein-scalar field equations for the class of
power law potentials $V(\phi) \propto \phi^{2n}$ are found in this
regime in terms of the inverse error function.

\vspace*{24pt}
\noindent
PACS number: 98.80.Cq

\noindent
\small Electronic Address: jim@fnas09.fnal.gov   \\
\phantom{Electronic Address:} iwaga@fnas10.fnal.gov

\end{abstract}

\normalsize

\end{titlepage}

%\double

\section{Introduction}

\setcounter{equation}{0}

\def\theequation{\arabic{equation}}

A study of the evolution of self-interacting scalar fields in the
early Universe is important for a number of reasons. Firstly, in
addition to offering a possible resolution to some of the fundamental
problems of the hot big bang model, the inflationary scenario provides
a causal mechanism for generating adiabatic density perturbations
\c{inflation}. These may produce anisotropies in the cosmic microwave
background and act as a seed for galaxy formation via gravitational
instability.  During inflation the Universe is dominated by the
potential energy $V(\phi)$ associated with the self-interactions of a
quantum scalar field $\phi$. If the field is initially displaced from
the global minimum of the potential and the potential is sufficiently
flat, the scalar field will evolve very slowly towards the true vacuum
state. The potential therefore behaves as an effective cosmological
constant and introduces a negative pressure into the Universe that
drives the accelerated expansion. Secondly, many natural extensions to
General Relativity may be expressed in terms of Einstein gravity
minimally coupled to a self-interacting scalar field after a suitable
conformal transformation on the metric tensor. Two classes of theory
that have received much attention in recent years are scalar-tensor
theories of gravity and higher-order theories, where the gravitational
lagrangian is an analytic function of the Ricci scalar $R$.

However, only a limited number of exact solutions to the isotropic
Einstein-scalar field equations have been found to date. These have
recently been classified by Barrow in terms of the potential $V=V_0
\phi^N \exp (A\phi^M)$, where $\{V_0,A,M,N\}$ are constants
\cite{barrow}. Special cases include $V={\rm constant}$ corresponding
to exponential expansion, power-law inflation from an exponential
potential $(N=0, M=1)$ \c{powerlaw,SB1990,SB1991,L1992} and
intermediate inflation from a combination of power-law potentials
$(N<0,M=0)$ \cite{Hphi,intermediate}. Solutions corresponding to
potentials leading to hybrid inflation \c{linde} have been presented
in Refs. \cite{EM1991,LIDSEY1} and exact solutions have also been
found for hyperbolic and trigonometric potentials
\cite{LIDSEY2,SC1992}.

 Consequently, it is common practice to invoke the {\em slow-roll}
approximation. This assumes that the kinetic energy of the scalar
field is negligible relative to its potential energy and that the
dominant term in the scalar field equation is due to friction arising
from the expansion of the Universe. The system is therefore reduced to
a set of coupled, first-order differential equations. The slow-roll
approximation is usually valid during the initial stages of the
inflationary expansion, but as the field rolls towards the minimum,
the approximation inevitably breaks down at some point. Moreover,
inflation arises whenever the strong energy condition is violated and
this does not necessarily require the slow-roll approximation to be
valid.

Therefore, it is important to develop analytical techniques that allow
deviations from the slow-roll regime to be accounted for and in this
paper we discuss how this may be achieved. After summarizing the
general features of scalar field dynamics in Sec. II, we illustrate
how corrections to the slow-roll approximation may be included in Sec.
III and we derive the corresponding field equations. In Sec. IV these
equations are solved in parametric form for the class of power law
potentials $V(\phi) \propto \phi^{2n}$ in terms of the inverse error
function. The rapid increase in the scale factor during inflation    is
naturally explained by using the properties of this function.

\section{Scalar Field Dynamics}

Considerable progress in determining scalar field dynamics has been
made recently by treating the scalar field as the dynamical variable,
since this reduces the field equations to a set of first-order,
non-linear differential equations
\c{Hphi,SB1990,SB1991,SC1992,Grish,lidseyplb}. The four-dimensional
action for Einstein gravity minimally coupled to a self-interacting
scalar field is
\be
\l{exponentialac}
S= \int d^4 x \sqrt{-g} \left[ \frac{R}{2\kappa^2} -\frac{1}{2} \left(
\nabla {\phi} \right)^2 -V( {\phi}) \right] ,
\ee
where $g\equiv {\rm det} g_{\mu\nu}$, $\kappa^2 \equiv 8\pi m_P^{-2}$
and $m_P$ is the Planck mass. We choose units such that $c=\hbar =1$.
If the Universe is spatially isotropic, closed and flat with a
world-interval $ds^2=-N^2(t) dt^2 +e^{2\alpha (t)} [dx^2+dy^2+dz^2]$
and lapse function $N(t)$, the Arnowitt-Deser-Misner (ADM) action is
\be
S=\int dt U N e^{3\alpha} \left[ -\frac{3}{\kappa^2}
\frac{\dot{\alpha}^2}{N^2} + \frac{1}{2} \frac{\dot{\phi}^2}{N^2}
-V(\phi) \right] ,
\ee
where $U\equiv \int d^3 {\bf x}$ is the comoving volume of the
Universe and a dot denotes differentiation with respect to cosmic time
$t$ \cite{ADM}. The classical Hamiltonian constraint ${\cal{H}}=0$
implies that the action satisfies the Hamilton-Jacobi equation
\be
\l{HJ}
 -\frac{\kappa^2}{6} \left( \frac{\partial S}{\partial \alpha}
\right)^2 + \left( \frac{\partial S}{\partial \phi} \right)^2 +2U^2
e^{6\alpha} V(\phi ) =0 .
\ee
The classical dynamics of this model is determined by the real,
separable solution
\be
\l{sepact}
S=-\frac{2}{\kappa^2} U e^{3\alpha} H(\phi) ,
\ee
where $H(\phi)$ satisfies the  differential equation \cite{Hphi,SB1990,SB1991}
\be
\l{SHJE}
\left( \frac{dH}{d\phi} \right)^2 -\frac{3\kappa^2}{2} H^2 (\phi) =-\frac{1}{2}
\kappa^4 V(\phi) .
\ee
This equation is equivalent to the $00$-component of the Einstein
field equations and therefore represents the Friedmann equation. In
the conformal gauge $N=1$ the momenta conjugate to $\alpha$ and $\phi$
are $p_{\alpha} = \partial S/\partial \alpha = -6\kappa^{-2}
Ue^{3\alpha} \dot{\alpha}$ and $p_{\phi}= \partial S/\partial \phi = U
e^{3\alpha} \dot{\phi}$, respectively. Substitution of ansatz
(\ref{sepact}) into these expressions implies that
\be
\l{field}
 H(\phi)=\dot{\alpha}, \qquad -\frac{2}{\kappa^2} \frac{dH}{d\phi}
=\dot{\phi}
\ee
and it follows that $H(\phi)$ is the Hubble expansion parameter
expressed as a function of the scalar field $\phi$. The general
solution to the field equations (\ref{SHJE}) and (\ref{field}) may be
expressed in terms of quadrature with respect to this function:
\be
\l{general}
\alpha [\phi (t)] = \alpha_i -\frac{\kappa^2}{2} \int^{\phi}_{\phi_i} d\phi' H
(\phi' ,p)\left( \frac{\partial H(\phi',p)}{\partial \phi'} \right)^{-1} ,
\ee
where $(\alpha_i,\phi_i)$ are arbitrary constants and $p$ is a
parameter associated with each solution $H(\phi ,p)$ of Eq.
(\ref{SHJE}) \cite{SB1990,SB1991}.

In principle the general path of the Universe in the minisuperspace
$(\alpha ,\phi )$ is uniquely determined once the functional form of
$H(\phi)$ is specified and this suggests that it should be be viewed
as the fundamental quantity in the analysis.  Unfortunately, however,
it is rather difficult to find exact solutions to Eq. (\ref{SHJE})
that go beyond the slow-roll approximation. In view of this it has
been suggested that one could begin the analysis by specifying
$H(\phi)$ \cite{LIDSEY1,L1993}. An alternative approach is to generate
exact solutions by viewing the expansion parameter as the effective
time coordinate \c{SM1994}.  The drawback with these approaches,
however, is that the exact solutions do not necessarily correspond to
realistic potentials. Moreover, since it is the potential, and not
$H(\phi)$, that is specified by the particle physics of the model, one
should aim to solve the model by specifying the potential.  In the
next section we show how an approximation to Eq. (\ref{SHJE}) can be
made when the slow-roll approximation is not valid.

\section{Beyond    Slow-Roll: The Andante Regime}

It  proves convenient to introduce the  rescaled quantities \c{Hphi}
\be
u\equiv \left( \frac{\kappa^2}{3}V\right)^{1/2}, \qquad x\equiv \left(
\frac{3}{2} \right)^{1/2} \kappa \phi
\ee
since    Eq. (\ref{SHJE}) then simplifies  to
\be
\l{rescaledenergy}
  H^2 =(H')^2+u^2 .
\ee
Deviations from the slow-roll regime have been analyzed previously by
Salopek and Bond \c{SB1990}. In order to derive a more accurate
solution they substituted the zeroth-order approximation $H_{(0)} =u$
into the right-hand side of Eq. (\ref{rescaledenergy}):
\be
\l{SBequation}
H^2_{(1)} = H^2_{(0)} \left[ 1+ \left( \frac{d\ln H_{(0)}}{dx}
\right)^2 \right]
\ee
and further accuracy is achieved by including higher-order terms.

 However, deviations from the slow-roll approximation may be studied
more quantitatively in terms of the dimensionless parameters
\bea
\epsilon \equiv  \frac{3\dot{x}^2}{ \dot{x}^2+9u^2}  = 3\left( \frac{H'}{H}
\right)^2  \nonumber \\
\eta \equiv -\frac{\ddot{x}}{H\dot{x}} = 3\frac{H''}{H} = \epsilon -\frac{1}{2}
\left( \frac{3}{\epsilon} \right)^{1/2} \epsilon'
\eea
and
\be
\xi \equiv 3\frac{H'''}{H'} =\eta - \left( \frac{3}{\epsilon} \right)^{1/2}
\eta' ,
\ee
where a prime denotes differentiation with respect to the
dimensionless field $x$ \cite{copeland}. For convenience we consider
$\dot{x}>0$ in this work, which implies that $\sqrt{\epsilon}
=-\sqrt{3}H'/H$. Modulo a constant of proportionality, $\epsilon$ is a
measure of the field's kinetic energy relative to its total energy,
whilst $\eta$ measures the field's acceleration relative to the amount
of friction acting on it due to the expansion of the Universe. We may
therefore refer to these quantities as the {\em energy} and {\em
friction} parameters, respectively \cite{LIDSEY2}.  The slow-roll
approximation to scalar field dynamics corresponds to $ \{|\epsilon |,
|\eta| , |\xi | \} \ll 1$ and in this regime the energy and friction
parameters reduce to the slow-roll parameters introduced in Ref.
\cite{slowroll}.  It is straightforward to show that inflation occurs
if $\epsilon <1$ and the end of inflation can be defined precisely by
the condition $\epsilon =1$ \cite{SB1990,SB1991,L1993}.

Eq. (\ref{rescaledenergy}) can be written in a very illuminating form
by introducing the parameter
\be
\l{v}
v  \equiv \sqrt{\frac{3}{\epsilon}}, \qquad v' = \frac{\eta}{\epsilon}-1 .
\ee
It follows that
\be
\l{definev}
\frac{1}{v^2} =\frac{\epsilon}{3} =1-\frac{u^2}{H^2}
\ee
and differentiation of this equation with respect to $x$ implies that
\be
\l{vequation}
\frac{u'}{u} = - \frac{1}{v} +\frac{v'}{v(v^2-1)} .
\ee
A comparison of Eqs.  (\ref{v}) and (\ref{vequation}) implies that the
`steepness' of the potential, as defined by Turner \c{turner}, may now be
expressed directly in terms of the energy and friction parameters:
\be
\l{steepness}
\frac{u'}{u} =-\sqrt{\frac{\epsilon}{3}}
\left[ \frac{1-\eta /3}{1-\epsilon /3} \right] .
\ee
The rate of change of steepness in the potential can also be written
with these parameters and has the form
\be
\l{changesteep}
\frac{1}{\kappa^2 V} \frac{d^2V}{d\phi^2}
=3\left[ \left( \frac{u'}{u} \right)^2 +\frac{u''}{u} \right]
=\frac{ \eta [1-\eta /3] + \epsilon [1-\xi /3]}{3[1-\epsilon /3]} .
\ee
Hence, the first slow-roll condition corresponds to $u'/u \approx
-\sqrt{\epsilon /3} \ll 1$, i.e. $H(x) \approx u(x)$.

We may consider first-order departures from the slow-roll regime by
expanding the terms in the square bracket of Eq. (\ref{steepness}) to
first-order in $\epsilon$ and $\eta$.  With the help of Eq. (\ref{v})
we find
\be
\l{andante}
\frac{u'}{u} \approx  -\frac{1}{v} +\frac{v'}{v^3}
+{\cal{O}} \left( \frac{v'}{v^5} \right)
\ee
and the dymanics of the scalar field in this regime of parameter space
is then determined by this equation. Eq. (\ref{andante}) may also be
derived by expanding the right-hand side of Eq. (\ref{vequation}) as a
geometric progression for $v>1$:
\be
\l{prog}
\frac{u'}{u} = - \frac{1}{v} +\frac{v'}{v^3}\left( 1+\frac{1}{v^2}
+ \frac{1}{v^4} + \ldots \right) .
\ee
It is seen, therefore, that accounting for first-order deviations from
the slow-roll regime is equivalent to including the first term in this
series.

For a specific choice of potential the solution to Eq. (\ref{andante})
will determine $v(x)$ and the functional form for $H(x)$ then follows
directly from Eq. (\ref{definev}). Eq. (\ref{andante}) describes the
dynamics of a scalar field that is evolving moderately slowly down its
self-interaction potential but at a sufficiently fast rate that the
usual slow-roll approximation is no longer valid. Consequently we
refer to this region as the moderately slow, or {\em andante}, regime
of scalar field dynamics\footnote{Similarly the slow-roll regime could
be referred to as the `largo' regime, whilst the epoch after inflation
that arises when the field undergoes rapid oscillations about the
minimum would represent the `allegro' regime.}. The slow-roll limit
corresponds to the regime where the energy and friction parameters are
negligible with respect to unity and the andante regime applies when
these parameters are small but finite. Expressions for the amplitudes
of the scalar and tensor fluctuations when these latter conditions
apply have been derived previously by Stewart and Lyth \cite{SL}.

An alternative form of Eq. (\ref{andante}) may be derived by viewing
the volume of the Universe as the effective dynamical variable. Since
the volume is a monotonically increasing function, it is a suitable
choice for the `time' coordinate. It follows from the general solution
(\ref{general}) that the number of $e$-foldings of expansion that
occur as the scalar field rolls from some initial value $x_i$ to a
value $x$ is given by
\bea
\l{efolds}
\ln a \propto N(x) =   \int^t_{t_i} dt' H(t')
= -\frac{1}{3} \int^x_{x_i} dx H(x')
\left( \frac{dH(x')}{dx'} \right)^{-1} \nonumber \\
= \frac{1}{3} \int^x_{x_i} dx' v(x') .
\eea
Hence, the ratio $z/z_i =(a/a_i)^3 = \exp (3N)$ represents the
fractional increase in the volume of the universe that occurs when the
field rolls from $x_i$ to $x$ and it follows immediately from the
definition (\ref{v}) that
\be
\l{z}
\frac{dx}{dz} =  \frac{1}{zv} .
\ee

If  we now define a new function $m(z)$:
\be
\l{mdefinition}
  m(z) \equiv  zx(z) ,
\ee
it can be shown, after differentiating with respect to $z$ and
substituting in Eq. (\ref{andante}), that
\be
\l{mveqn}
\frac{dm}{dz} = x+\frac{1}{v}
\ee
and
\be
\l{mequation}
m\frac{d^2 m}{dz^2} =-x\frac{u'}{u} .
\ee
We shall show in the following section that Eq. (\ref{mequation}) is
exactly solvable when the potential is given by $V(\phi) \propto
\phi^{2n}$.

\section{Power Law Potentials}

In this section we consider the class of power law potentials
\be
\l{polynomial}
V(\phi) =   \lambda    \phi^{2n}, \qquad n={\rm constant} .
\ee
where ${\lambda}$ is the coupling constant. In the chaotic
inflationary scenario, the andante regime should apply during the
final stages of the inflationary expansion when the scalar field
approaches the global minimum of its potential.  In this region it is
useful to consider the general inflationary potential as a truncated
Taylor series expanded about this minimum. It is therefore a good
approximation to assume the potential has the power law form
(\ref{polynomial}) with positive-definite $n$. However, the class of
potentials with $n<0$ is also interesting and potentials of this form
may be important if the Universe contains a nonvanishing vacuum energy
at the present epoch \c{PR}. Moreover, they arise in generalized
scalar-tensor theories when the Brans-Dicke parameter is viewed as a
truncated Taylor series in the dilaton field \c{BM1990}.

When  the potential is a power law,
Eq. (\ref{mequation}) admits  the first integral
\be
\l{firstintegral}
  \frac{dm}{dz}   =   \pm \left( c-2n\ln m  \right)^{1/2} ,
\ee
where $c$ is an arbitrary constant. For consistency we require $m>0$
in Eq. (\ref{firstintegral}), and since we assumed $\dot{x}>0$, this
implies that both $x$ and $z$ must be negative if $n>0$ and they must
be positive if $n<0$. However, $|z|$ provides a measure of the amount
of inflation that occurs and it varies much more rapidly than the
scalar field $x$. Thus, the evolution of the function $m$ is dominated
by $z$, which implies that $m$ increases as $|z|$ increases.
Consequently, the negative square root should be chosen when $n>0$ and
the positive square root corresponds $n<0$ .

Eq. (\ref{firstintegral}) is solved exactly in terms of the error
function:
\be
\l{zsolution}
z=b+\gamma {\rm erf} \left[ \mp \left( \frac{c}{2n} -\ln ( zx)
\right)^{1/2} \right] ,
\ee
where $b$ is the second integration constant and
\be
\gamma \equiv \left( \frac{\pi}{2n} \right)^{1/2} e^{c/(2n)} .
\ee
This solution may be derived from the identity $d[{\rm erf}(y)]/dy
=(2/\sqrt{\pi})\exp (-y^2)$ \cite{AS}.

Eq. (\ref{zsolution}) can be inverted to yield
\be
\l{xsolution}
x(z) = e^{c/(2n)} \frac{1}{z} \exp \left[ - \left( {\rm Ierf} \left(
\frac{z-b}{\gamma} \right) \right)^2 \right] ,
\ee
where ${\rm Ierf}(y)$ is the inverse error function and the expression
for $v(z)$ follows by substituting Eqs.  (\ref{firstintegral}) and
(\ref{xsolution}) into Eq. (\ref{mveqn}):
\be
\l{vsolution}
\frac{1}{v} =-x-\sqrt{2n} {\rm Ierf} \left( \frac{z-b}{\gamma} \right) .
\ee
Finally we find $z(x)$ by substituting Eq. (\ref{vsolution}) into Eq.
(\ref{xsolution}):
\be
\l{z(x)solution}
z(x) = e^{c/(2n)} \frac{1}{x} \exp \left[ -\frac{1}{2n} \left(
x+\frac{1}{v(x)} \right)^2 \right] .
\ee

These solutions exhibit the correct qualitative behaviour that one
would expect for a scalar field rolling down a polynomial potential.
Figures 1a and 1b illustrate the behaviour of the functions $x(z)$ and
$v^{-1}(z)$ for $n=1$, $c=0$ and $b=-\gamma$ and similar results are
found for others choices of these constants.  A relatively small
change in the value of the scalar field in the range $-30 \le x \le
-10$ results in a huge change in the value of $z$ by a factor $\sim
10^{180}$. The origin of this behaviour is traced to the properties of
the inverse error function. ${\rm Ierf}(y)$ is an odd function of $y$
and is undefined for $|y|>1$. For positive arguments it is a
monotonically increasing function with ${\rm Ierf}(0)=0$ and ${\rm
Ierf}(1)=\infty$. However, the slightest deviation of the argument
from unity results in a sharp decrease in the value of the inverse
error function. For example, ${\rm Ierf}(1-10^{-1000})=47.9$ and ${\rm
Ierf}(1-10^{-100})=15.1$. This implies that ${\rm Ierf}(y)$ is a
relatively slowly varying function in the range $0\le y\le
1-10^{-10^3}$ and consequently $z$ may change by a factor $10^{10^3}$
without there being a significant change in the value of the scalar
field. This feature leads to a rapid growth in the scale factor for a
very small change in the value of the scalar field.

In the limit $|x| \gg v^{-1}$ the functional form of the solution
(\ref{z(x)solution}) implies that the scale factor may be expressed
directly in terms of the scalar field as
\be
a(x) \propto \left( \frac{1}{x} \right)^{1/3} e^{-x^2/(6n)} .
\ee
{}From this expression it follows that Eq. (\ref{efolds}) may be
employed to evaluate $H(x)$:
\be
H^2=u^2 \left[ 1+\frac{n}{x^2} \right]^n ,
\ee
and if one expands the right-hand side to first-order, the result is
equivalent to Eq. (\ref{SBequation}). Hence, the Salopek-Bond
approximation \c{SB1990} is recovered from the parametric solution
(\ref{xsolution})-(\ref{vsolution}) in the limit that $v$ diverges.

Similar arguments apply when $n<0$. Considering negative values of $n$
is equivalent to Wick rotating $\gamma$ to the imaginary axis, i.e.
$\gamma \rightarrow i\gamma$. The error function with imaginary
argument is related to the imaginary error function, ${\rm erf}(iy)
\equiv i {\rm erfi} (y) $, and it follows that solutions
(\ref{xsolution}) and (\ref{vsolution}) take the form
\be
\l{1}
x(z) = e^{c/(2n)} \frac{1}{z} \exp \left[ \left( {\rm Ierfi} \left(
\frac{z-b}{\sqrt{|\gamma^2|}} \right) \right)^2 \right]
\ee
and
\be
\l{2}
\frac{1}{v} =-x +\sqrt{2|n|} {\rm Ierfi}
\left( \frac{z-b}{\sqrt{|\gamma^2|}} \right) ,
\ee
respectively.  Figures 2a and 2b illustrate the evolution of $x(z)$
and $v^{-1}(z)$.  The De Sitter solution is the attractor at infinity
for these models and the intermediate inflationary solution is
recovered at large $x$ \c{intermediate}. Solutions (\ref{1}) and
(\ref{2}) illustrate analytically how these asymptotic solutions are
reached.

\section{Conclusion}

In this paper we have considered deviations from the slow-roll
approximation by including the first-order contributions from the
kinetic energy and acceleration of the field. This allows a more
accurate analysis of scalar field dynamics to be performed. Such
improvements are expected to be relevant towards the final stages of
inflation when the scalar field is close to the global minimum of its
potential. In view of this we considered a class of power law
potentials and an exact parametric solution to the field equations was
found in terms of the inverse error function. This function exhibits
some very interesting properties and it was shown how the rapid growth
in the scale factor of the Universe during inflation is naturally
explained by employing the properties of this function. Since a large
increase in the scale factor for a very small change in the value of
the scalar field is a generic feature of the inflationary scenario, we
conjecture that the inverse error function may arise in the general
solution to the field equations. This possibility is currently under
investigation. The solutions presented here should improve our
understanding of how the Universe moves out of the inflationary epoch
and into the reheating phase.  It will be interesting to investigate
whether solutions to Eqs. (\ref{andante}) and (\ref{mequation}) can be
found for other potentials.

\vspace{.1in}

{\em Note}: After this work was completed we received a preprint by
Liddle {\em et al.} that also discusses the slow-roll approximation
though with a different method of analysis \c{LPB}.

\vspace{.1in}

\centerline{\bf Acknowledgements}

\vspace{.1in}

We would like to thank Luca Amendola, Robert Caldwell and Josh Frieman
and Andrew Liddle for helpful conversations. JEL is supported by the
Particle Physics and Astronomy Research Council (PPARC), UK. IW is
supported in part by the Brazilian agency CNPq. The authors are
supported at Fermilab by the DOE and NASA under Grant No. NAGW-2381.

\vspace{.1in}

%%%%%%%%%%%%%%%%%%%%%%%
\frenchspacing
\def\prl#1#2#3{{ Phys. Rev. Lett.} {\bf #1}, #2 (#3)}
\def\prd#1#2#3{{ Phys. Rev. D} {\bf #1}, #2 (#3)}
\def\plb#1#2#3{{ Phys. Lett. B} {\bf #1}, #2 (#3)}
\def\npb#1#2#3{{ Nucl. Phys. B} {\bf #1}, #2 (#3)}
\def\apj#1#2#3{{ Ap. J.} {\bf #1}, #2 (#3)}
\def\apjl#1#2#3{{ Ap.J. Lett.} {\bf #1}, #2 (#3)}
\def\cqg#1#2#3{{Class. Quantum Grav.} {\bf #1}, #2 (#3)}
\def\grg#1#2#3{{Gen. Rel. Grav.} {\bf #1}, #2 (#3)}
\def\mnras#1#2#3{{Mon. Not. R. astron. Soc.} {\bf #1}, #2 (#3)}
%%%%%%%%%%%%%%%%%%%%%%%%%%%%%%

\newpage

\centerline{\bf Figure Captions}

\vspace{1in}

{\em Figure 1}: (a) A plot of the solution (\ref{xsolution})
representing the evolution of the scalar field $x$ with respect to
$-{\rm log}_{10} z$ for the quadratic potential $u(x)\propto x$
$(n=1)$. The numerical values of the integration constants are
specified to be $c=0$ and $b=-(\pi/2)^{1/2}$.  (b) The solution
(\ref{vsolution}) for the same choice of parameters as in Figure 1a.

\vspace{1in}

{\em Figure 2}: (a) Illustrating the solution (\ref{1}) for the
decaying power law potential $u\propto x^{-1}$ $(n=-1)$, where $b=c=0$
and $\gamma=(\pi/2)^{1/2}$. (b) The corresponding solution (\ref{2}).
The solutions approach the intermediate inflationary solution and the
De Sitter solution is the attractor at infinity. Hence, there is no
exit from inflation for these potentials.


\centerline{{\bf References}}
\begin{enumerate}
\bibitem{inflation}
A. H. Guth and S.-Y. Pi, \prl{49}{1110}{1982}; \\
S. W. Hawking, \plb{115}{295}{1982}; \\
A. A. Starobinskii, \plb{117}{175}{1982}.


\bibitem{barrow}
J. D. Barrow, \prd{48}{1585}{1993}.

\bibitem{powerlaw}
F. Lucchin and S. Matarrese, \prd{32}{1316}{1985}; \\
J. D. Barrow, A. B. Burd and D. Lancaster, \cqg{3}{551}{1986}; \\
J. Halliwell, \plb{185}{341}{1987}; \\
J. D. Barrow, \plb{187}{341}{1987}; \\
A. B. Burd and J. D. Barrow, \npb{308}{929}{1988}; \\
A. R. Liddle, \plb{220}{502}{1989}. \\


\bibitem{Hphi} A. G. Muslimov, \cqg{7}{231}{1990}.


\bibitem{SB1990}
D. S. Salopek and J. R. Bond, \prd{42}{3936}{1990}.

\bibitem{SB1991}
D. S. Salopek and J. R. Bond,  \prd{43}{1005}{1991}.

\bibitem{L1992}
J. E. Lidsey, \cqg{9}{1239}{1992}.

\bibitem{intermediate}
J. D. Barrow, \plb{235}{40}{1990}; \\
J. D. Barrow and P. Saich, \plb{249}{406}{1990}; \\
J. D. Barrow and A. R. Liddle, \prd{47}{R5219}{1993}.



\bibitem{linde} A. D. Linde, \prd{49}{748}{1994}.


\bibitem{EM1991} G. F. R. Ellis and M. S. Madsen, \cqg{8}{667}{1991}.

\bibitem{LIDSEY1}
J. E. Lidsey, \cqg{8}{923}{1991}.


\bibitem{LIDSEY2}
B. J. Carr and J. E. Lidsey, \prd{48}{543}{1993}; \\
J. E. Lidsey, \mnras{266}{489}{1994}.

\bibitem{SC1992}
S. P. Starkovich and F. I.  Cooperstock, \apj{398}{1}{1992} \\
J. D. Barrow, \prd{49}{3055}{1994}.



\bibitem{Grish}
L. P. Grishchuk, \cqg{10}{2449}{1993}.

\bibitem{lidseyplb}
J. E. Lidsey, \plb{273}{42}{1991}.



\bibitem{ADM} R. Arnowitt, S. Deser and C. W. Misner, in {\em Gravitation: An
Introduction to Current Research}, edited by L. Witten (John Wiley, New York,
1962); \\
M. A. H. MacCallum, in {\em Quantum Gravity: An Oxford Symposium},
edited by C. J. Isham, R. Penrose and D. W. Sciama (Oxford University
press, Oxford, 1975).


\bibitem{L1993}
J. E. Lidsey, \grg{25}{399}{1993}.



\bibitem{SM1994}
F. E. Schunck and E. W. Mielke, "A New method for Generating Exact
Inflationary Solutions," Preprint gr-qc/9407041, 1994, unpublished.


\bibitem{copeland}
E. J. Copeland, E. W. Kolb, A. R. Liddle and J. E. Lidsey,
\prd{49}{1840}{1994}.

\bibitem{slowroll}
P. J. Steinhardt and M. S. Turner, \prd{29}{2162}{1984}; \\
A. R. Liddle and D. H. Lyth, \plb{291}{391}{1992}.




\bibitem{turner}
M. S. Turner, \prd{48}{3502}{1993}.

\bibitem{SL} E. D. Stewart and D. H. Lyth,  \plb{302}{171}{1993}.

\bibitem{PR} B. Ratra and P. J. E. Peebles, \prd{37}{3406}{1988}.

\bibitem{BM1990} J. D. Barrow and K. Maeda, \npb{341}{294}{1990}.


\bibitem{AS}
{\em Handbook of Mathematical Functions},
edited by M. Abramowitz and I. A. Stegun, Natl. Bur.
Stand. Appl. Math. Ser. No. 55 (U.S. GPO, Washington, D.C., 1965).

\bibitem{LPB} A. R. Liddle, P. Parsons and J. D. Barrow, "Formalising the
Slow-Roll Approximation in Inflation", Sussex Preprint (1994).

\end{enumerate}
\end{document}